\documentclass[12pt,a4paper]{article}
\usepackage[T2A]{fontenc}
\usepackage[cp1251]{inputenc}
\usepackage[english]{babel}
\usepackage{amssymb}
\usepackage{amsmath}
\usepackage{graphicx}
\usepackage[small,sc,center]{titlesec}
\usepackage{indentfirst}
\usepackage[labelsep=period]{caption}
\usepackage{caption}
\newcommand{\msun}{\,$M_{\odot}$}

\newcommand{\rsun}{\,$R_{\odot}$}
\newcommand{\ergs}{\,erg\,s$^{-1}$}
\newcommand{\gcmq}{\,g\,cm$^{-3}$}
\newcommand{\gcm}{\,g\,cm$^{-1}$}
\newcommand{\kms}{\,km\,s$^{-1}$}
\newcommand{\cms}{\,cm\,s$^{-1}$}

\newcommand{\cmsqg}{\,cm$^2$\,g$^{-1}$}
\newcommand{\ha}{H$\alpha$}
\begin{document}
	
\begin{center}
\textbf{\large Circumstellar shell and presupernova emission of SN~2020tlf} 

\vskip 5mm
\copyright\quad
2022 г. \quad N. N. Chugai\footnote{email: nchugai@inasan.ru} and V. P. Utrobin$^{2,1}$\\
\textit{$^1$Institute of astronomy, RAS, Moscow} \\
\textit{$^2$NIC ''Kurchatov Institute``, Moscow} \\

Submitted  12.04.2022 г.
\end{center}


\vspace{1.2cm}
 
 \begin{abstract} 
We address a phenomenon of a confined circumstellar (CS) dense shell and powerful presupernova emission 
 of SN~2020tlf (type IIP). 
Modeling the \ha\ line and the circumstellar interaction implies the CS 
  shell radius of $\sim$10$^{15}$\,cm and the mass of $\sim$0.2\msun\ lost during 
  $\sim$6 yr prior to the explosion.
Spectra and photometry of the supernova after the explosion do not show apparent 
  signature of the material lost by the presupernova during its powerful luminosity. 
This material presumably resided in the inner zone of the CS shell.    
We present a hydrodynamic model of the outcome of a flash with the energy of  
  $5\times10^{48}$\,erg in the convective nuclear burning zone.
The model predicts the ejection of outer layers of the presupernova ($\sim$0.1\msun)  and 
  the luminosity of  $10^{40}$\ergs\ during several hundreds days in accord with 
    observations.
We propose the Lighthill mechanism of acoustic waves generation by the  
  turbulence of the convective nuclear burning zone to account for the phenomenon 
  of a compact CS shell of supernovae related to the core collapse.

\end{abstract}

\section{Introduction}

Last 20 years the study of core collapse supernovae (CCSN) have revealed  
 a challenging problem of the vigorous mass loss by presupernova (preSN) year--decade prior 
  to the explosion.
This phenomenon undoubtedly related somehow to the enhanced rate of thermonuclear 
 burning at the final stage of massive star, albeit details of this connection 
  are far from clear.  
Signes of the vigorous mass loss are manifested in early spectra as emission 
lines of specific profiles (narrow core and broad wings) of 
  hydrogen, if any, He\,II and N\,III. 
That sort of lines were observed in early spectra of type IIL SN~1998S (Fassia et al. 2001)  
 and recognized as emission of a compact ($\sim$10$^{15}$\,cm) 
 dense CS shell with a large Thomson optical depth (Chugai 2001).
Similar compact CS shells are found in SNe~IIP (Yaron et al. 2017)   and SNe~Ibn 
 (Pastorello et al. 2015).
 
In this row is the type II SN~2020tlf that should be classified as IIP, since 
  the light curve has a plateau with a clear-cut transition to the radioactive tail.
Apart from a confined dense shell, this supernova is remarkable for 
  the preSN emission during 130 days before the explosion with the super-Eddington luminosity 
  of $\approx$10$^{40}$\ergs\ that presumably is associated with the formation of the CS shell 
  (Jacobson-Gal\'{a}n et al. 2022).
Authors conclude that it is the matter lost at the stage of preSN high luminosity that 
 is observed on day 10 in CS emission lines.   
Modeling the light curve and spectra implies that the CS shell is produced by the 
 preSN mass loss rate of 0.01\msun\,yr$^{-1}$ and limited to $r \approx 10^{15}$\,cm
   (Jacobson-Gal\'{a}n et al. 2022).    
  
This scenario however requires some revision.
The point is that hydrodynamic modeling of the shell ejection initiated by the 
 $5\times10^{48}$\,erg deposition in the preSN core suggests that 
 the ejecta expands at $\sim$50\kms\ (Dessart 2010).
Combined with the preSN radius ($R_0 \sim 500$\rsun) the ejected shell expands up to 
$r \sim R_0$ + (130 d) $\times$ (50\kms) $\sim10^{14}$\,cm.
This CS gas after the SN explosion should be swept up by the 
  forward shock with the velocity of $\sim$10$^4$\kms\ 
  on the time scale of one day.
On the other hand, CS emission lines were observed in the spectrum on day 10 
 after the explosion, which implies that the CS line-emitting gas is not related to the mass lost 
  during preSN high-luminosity stage and the issue of the structure and mass of the CS shell remains open.
   
A central issue, however, is the mechanism of powerful preSN emission during 130 days and 
 the amount of matter lost during this period.   
One explanation of the enhanced mass loss before the supernova explosion is based on a  
 generation of gravitational hydrodynamic waves caused by the vigorous convection in 
 the burning zone of C, O, and Ne.
Gravitational waves presumably convert into the acoustic waves that dissipate in the 
 preSN envelope thus causing the enhanced mass loss (Quataert and Shiode 2012, Leung et al. 2021).
An alternative scenario (Jacobson-Gal\'{a}n et al. 2022) suggests that a rapid release
 of the nuclear energy in the internal zone, comparable to the binding energy of
 the external envelope, drives the shock wave and the subsequent matter ejection. 
The mechanism is studied earlier and it has been shown that the shell ejection is accompanied 
  by a prolong powerful luminosity during several hundreds days (Dessart 2010).
The mechanism however has not been explored in the particular case of SN~2020tlf.      
 
Our paper pursues two goals: (i) the determination of structure and parameters of the CS envelope
 based on the modeling of the \ha\ emission line on day 10 in combination with the modeling of the light curve 
 powered by the CS interaction; (ii) modeling the preSN mass ejection and prolong luminosity initiated 
 by the energy release in nuclear burning zone. 
In line with the formulated goals we describe modeling methods in section 2, report results in section 3, and 
  discuss them in the final section.

\section{Models used}

Data analysis and inferences are based on the application of three types of models: 
 (i) the model for the \ha\ line formed in the expanding CS 
 envelope with the large Thomson optical depth; 
 (ii) the model of the supernova interaction with the CS matter; and (iii) the hydrodynamic 
 model of mass ejection and emission initiated by the energy deposition in the central zone of 
 preSN.

\subsection{Two components of CS envelope}

Previous modeling of the bolometric light curve (Jacobson-Gal\'{a}n et al. 2022) suggests that 
  the preSN of SN~2020tlf was a red supergiant (RSG) with the initial mass of about 12\msun\ and 
  radius of $R_0 \sim 500-1100$\rsun.
The matter lost by the preSN during the powerful luminosity (130 days) 
  occupied a region of $r \lesssim r_1 = R_0 + (130\,\mbox{days})\times(50\,\mbox{\kms}) \sim 10^{14}$\,cm
  at the supernova explosion,
  whereas the gas responsible for the emission lines on day 10 occupied a zone  
 $r \lesssim r_2 = (10^4\,\mbox{\kms})\times(10\,\mbox{days}) \sim 10^{15}$\,cm.
 
These estimates indicate that the CS envelope of SN~2020tlf has two-component structure characterized by 
  the presence of a ''core`` ($r \lesssim r_1 \sim 10^{14}$\,cm) related to the matter lost during the powerful preSN luminosity and 
  ''halo``  ($r \lesssim r_2 \sim 10^{15}$\,cm) identified with the wind lost before the preSN flash. 
 In our light curve model the core has uniform density within $r < r_1$, whereas the halo 
  density distribution ($r_1 < r < r_2$) corresponds to the steady wind $\rho = Ar^{-2}$.
In fact, the density distribution in the core observationally is of little significance since this matter is swept up by 
the supernova in $1-2$ days after the explosion.

\subsection{\ha\ model}

The \ha\ emission on day 10 originates from the preshock CS gas. 
Due to a strong radiative cooling the forward shock wave almost coincides with the 
  cold dense shell (CDS) that forms between the forward and reverse shock. 
At this stage the CDS optical depth is large, so the photosphere essentially 
 coincides with the CDS.
In the adopted model the region of the \ha\ formation is the wind ($\rho = Ar^{-2}$) 
  between the photosphere and the external radius ($r_p < r <r_2$); the CDS radius is 
  provided by the CS interaction model.
    
At early stage the supernova radiation significantly accelerates preshock gas, so that the 
  initial wind velocity ($u_0 \approx 50$\kms) becomes larger by the amount related to the 
  acceleration by the supernova radiation
\begin{equation}
u_{ac} = \frac{kE_r}{4\pi r^2c} = 90E_{r,49}r_{15}^{-2}\quad \mbox{\kms}\,,  
\label{eq:accel}  
\end{equation}  
where $k = 0.34$\cmsqg\ is opacity and $c$ is the speed of light. 
The preshock velocity $u_{ac}$ is provided by the CS interaction model.

Modeling the \ha\ spectrum is based on the Monte Carlo technique.
This suggests photon history dicing starting from a emission within the thermal line profile.
The photon may leave the CS shell after multiple Thomson scatterings off thermal electrons or 
 may vanish colliding with the photosphere with the probability $(1-\Omega$), where $\Omega$ is 
 the photosphere albedo.
During a random walk in the CS shell the photon frequency changes 
 in the comoving frame and also due to the frequency redistribution at the scattering off thermal electrons.
The latter is diced using the angle-averaged frequency redistribution function (Hummer \& Mihalas 1967)   
 with the comptonization correction (energy exchange between photons and electrons).
The electron temperature is assumed to be constant along the radius.   
The optimal spectral model (SM) and hydrodynamic model (HM) of the CS interaction 
 are the outcome of the iterative process HM $\rightarrow$ SM  $\rightarrow$ HM $\rightarrow$ SM ...

\begin{table}[t]
	\vspace{6mm}
	\centering
	{{\bf Table 1} \ha\ model parameters. Indicated in parentheses is the power of ten. }
	
	\vspace{5mm}\begin{tabular}{c|c|c|c} 
	\hline	
	  &     &    &   \\
		Model  & $A$ (g/cm)  & $r_2$~(cm) & $T_e$ (K)  \\
		\hline
		A      &   3.2(16)    &  1.2(15)   & 10000    \\  
		B      &   3.2(16)    &  0.8(15)   & 10000    \\  
		C      &   1.6(16)    &  1.2(15)   & 10000  \\ 
		D      &   3.2(16)    &  1.2(15)   & 20000   \\ 
		\hline
	\end{tabular}
  
\end{table}

\subsection{Supernova interaction with CS gas}

Hydrodynamics of the CS interaction is described in the thin shell approximation 
 (Guiliani 1981, Chevalier 1982) with a correction on the forward shock acceleration 
  in the region of steep density drop ($\rho \propto r^{-\omega}$, $r > r_2$, $\omega > 3$) 
  following transition to the adiabatic regime.
The similar model has recently been used (Chugai 2022), so we bound ourselves with its short description.

The kinetic luminosity of the forward and reverse shock converts into X-rays which 
  are absorbed by the unperturbed ejecta, the CDS, and the CS gas and are then re-emitted as the observed 
  optical radiation.
The X-ray luminosity of a shock wave at the moment $t$ is the kinetic luminosity multiplied by the 
  radiation efficiency $\eta = t/(t+t_c)$, where $t_c$ is the cooling time of the shocked 
    hot gas with the cooling function of Sutherland \& Dopita (1993).
The absorbed fraction of the X-ray luminosity is calculated assuming the
   bremsstrahlung spectrum and the absorption coefficient
   $k_X = 100(h\nu/1\,\mbox{keV})^{-8/3}$\,cm$^2$\,g$^{-1}$.
   
\begin{figure}
	\centering
	\includegraphics[trim = 20 120 0 40, width=0.97\textwidth]{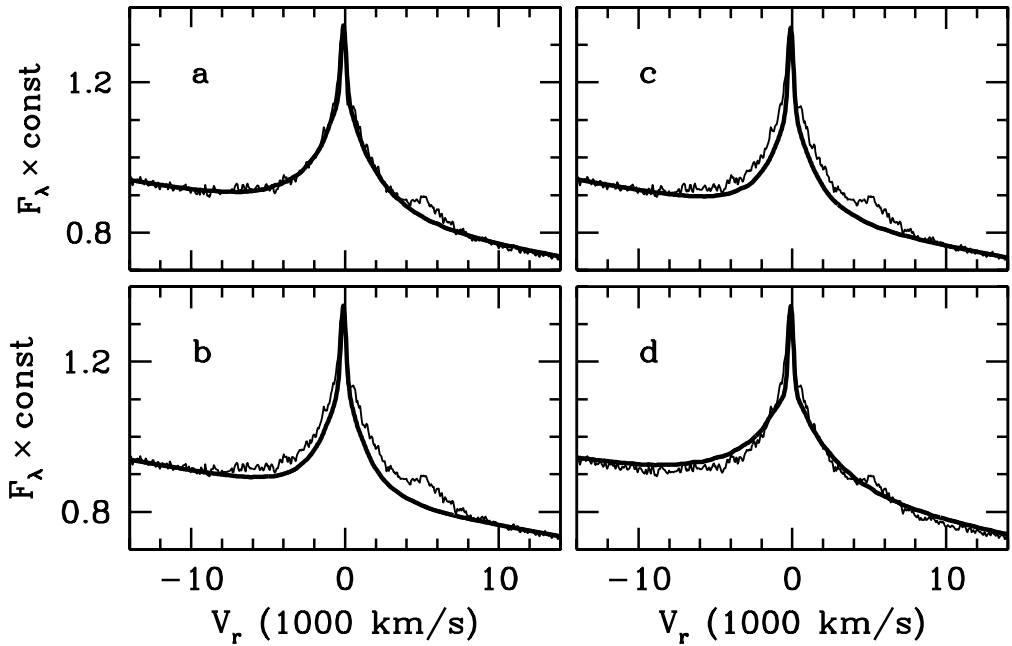}
	\caption{\rm
		Model \ha\ spectrum compared to the observed one ({\em thin line}). 
		Panels a, b, c, and d show models A, B, C, and D, respectively (Table 1).
		Emission line around  +5000\kms\ is He\,I 6678\AA\ that is not included in the model.
	}
	\label{fig:blend}
\end{figure}

The model bolometric luminosity at the moment $t$ suggests the instant re-emission of the 
  absorbed X-rays into the escaping optical radiation.
However at the initial stage of the order of the diffusion time for the photons in 
 the CS shell --- defined by the condition $t_{dif} = t_{esc} = t$, where 
  $t_{esc} = 0.5\tau r/c$ is the 
 average escape time from the homogeneous sphere of the optical depth $\tau$ with 
  the central source (Sunyaev \& Titarchuk 1980) --- the corresponding diffusion delay should be taken 
  into account.     
This is done according to a simplified recipe: the luminosity calculated without 
 diffusion is multiplied by a smoothed step function $S(x) = x^9/(1 + x^9)$, where 
 $x = t/t_{dif}$.
The model bolometric light curve is the superposition of the luminosity due to the 
 CS interaction and the diffusion supernova luminosity calculated according to the 
  analytical model (Arnett 1980). 
  
The model of the CS interaction is aimed at the description of the initial stage 
  of the light curve and the maximal velocity of the undisturbed supernova envelope. 
The lower limit of this velocity is estimated from the maximal radial velocity 
 in the blue wing of Ca\,II 8498\AA\ absorption line on day 95  (Jacobson-Gal\'{a}n et al. 2022).  
For a given CS gas density the light curve and the maximal velocity supernova 
  before the reverse shock depend on the kinetic energy and the mass of the supernova ejecta. 
We therefore fix the energy at the maximal level of $2\times10^{51}$\,erg for the neutrino-driven mechanism 
 (Janka 2017), whereas the mass is considered as a free parameter. 
The initial supernova model suggests homologous expansion with density $\rho = \rho_0/[1 + (v/v_0)^n]$, 
where $v_0$ and $\rho_0$ are determined by the ejecta mass 
and the kinetic energy assuming   $n = 8$.

  \subsection{Modeling preSN mass ejection and luminosity}
    
The underlying model for the preSN is a star with the initial mass of 12\msun\ in hydrostatic equilibrium 
  at the final stage (Woosley et al. 2002). 
This choice is in line with the earlier conclusion that the SN~2020tlf progenitor was a star with the 
  initial mass of $10-12$\msun\  (Jacobson-Gal\'{a}n et al. 2022). 
  
Based on the radiation hydrodynamics code \textsf{CRAB} (Utrobin 2007), we compute the 
 evolution of a perturbation initiated by the instant release of an internal energy  $E_{dep}$ 
  at the level with a mass coordinate $m_{dep}$. 
In line with the Dessart et al. (2010) modeling of an outcome of the energy release at the level  $m_{dep} = 1.8$\msun\ 
 we ignore details of a thermonuclear flash of C, O, Ne burning and remark only that the burning of 
  0.01\msun\ of C or/and O is enough to release $E_{dep} \approx 2\times10^{49}$\,erg.
As becomes clear below, the value of $E_{dep}$ should be a significant fraction  ($q$) of the 
  binding energy $E_b$ of the matter lying above in order to reproduce the observed preSN luminosity by this hydrodynamic perturbation.

\begin{table}[t]
	\centering
	{{\bf Table 2.} Parameters of CS shell. For the radius the number in parentheses is the power of ten.}
	
	\vspace{6mm} \begin{tabular}{l|c|c}
		\hline
		Parameter & Core  &  Halo \\
		\hline
		&      & \\
		Radius (cm)  & 2.7(14)  & 1.2(15)   \\	
		Mass (\msun)   &  0.14  & 0.22 \\
		Expansion velocity (\kms)  & 45 & 50 \\
	\hline	
	\end{tabular} 	
\end{table}

\section{ Modeling results} 

\subsection{Parameters of CS shell halo}

Key parameter for the \ha\ spectral model is the Thomson optical depth of 
  the CS shell.
Instead of the optical depth, we use, however, the external 
  radius $r_2$ that is determined by the width of the initial luminosity peak 
  and the wind density parameter $A$.
To reproduce the light curve and the \ha\ line requires
 $r_2 = 1.2\times10^{15}$ cm.
On day 10, when the spectrum was obtained, the CDS (i.e., photosphere) radius 
  in the CS interaction model is $r_p = 6.4\times10^{14}$\,cm. 
The photosphere is assumed to be fully absorbing ($\Omega = 0$).   

Four \ha\ models (Figure 1) are distinguished by the density, the radius $r_2$, and 
  the electron temperature ($T_e$) (Table 1).
The CS expansion velocity is the sum $u = u_0 + u_{ac}$ with the initial wind velocity 
 $u_0 = 50$\kms.
On day 10 the velocity $u_{ac}$ produced by the preshock acceleration is 270\kms\ at the 
 radius $r_p$ and 80\kms\ at the radius $r_2$.
Between these extreme values the velocity $u_{ac}$ is interpolated linearly.   
The standard model A (Table 1) produces qualitatively best fit.
The model B with a smaller halo radius manifests the effect of a slightly lower optical depth.
Models C and D show the pronounced effects of the lower density and the higher electron temperature.
The major parameters of the halo of the CS shell are presented in Table 2.

Note that models are compared to the observed spectrum in the rest frame assuming supernova redshift 
2369\kms\ that is by 144\kms\ lower than the host galaxy NGC 5731 redshift 2513\kms\ (database NED).
The different redshift for the supernova and the galaxy presumably is related to the 
the galactic rotation, since the galaxy is seen almost edge-on (inclination $i =82^{\circ}$, NED) and 
the displacement from the galacic center is significant ($9''$).
Unfortunately this explanation cannot be confirmed as yet because of the absence of data on the  NGC 5731 
rotation curve.

\begin{figure}
	\centering
	\includegraphics[trim = 50 120 0 0, width=0.97\textwidth]{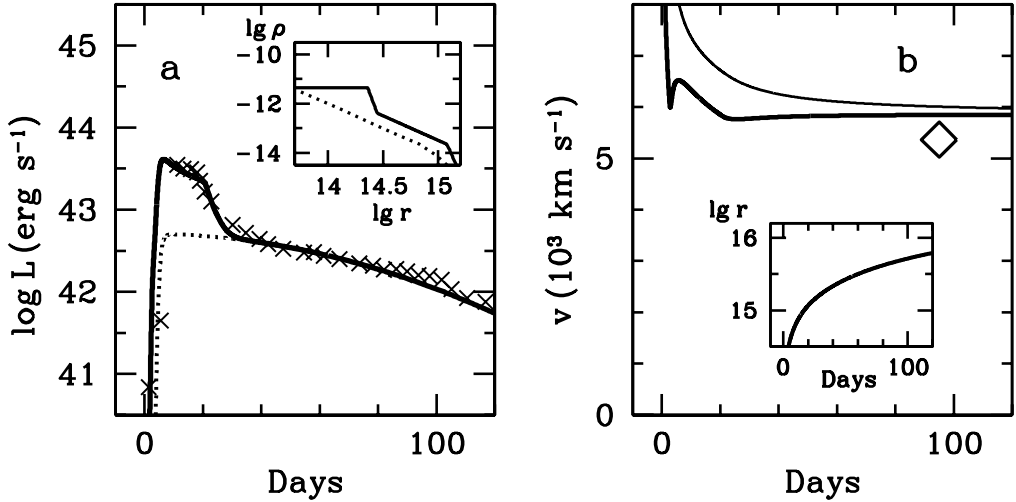}
	\caption{\rm
		{\em Left.} Model bolometric light curve 
		compared to the observational one ({\em crosses}). 
		The supernova diffusion luminosity without the CS interaction is shown by {\em dotted line}.
		{\em Inset} shows the density distribution in the CS shell compared to 
		the distribution from Jacobson-G\'{a}lan et al. (2022) ({\em dotted line}).\\
		{\em Right.} Model CDS velocity ({\em thick line}) and the boundary velocity of the unperturbed 
		supernova envelope ({\em thin line}). The maximal velocity obtained from the Ca\,II 8498\AA\ line on 
		day 95 is shown by {\em  diamond}. {\em Inset} shows the evolution of the CDS radius.  
	}
	\label{fig:blc1}
\end{figure}

Parameters of the CS shell halo are used for the CS interaction model.
As to the mass and the radius of the CS shell core, they are supplied by the 
 hydrodynamic model of the preSN luminosity (see below).
Two models are presented.
The standard model (Figure 2) is characterized by the ejecta mass of 9\msun, 
  the kinetic energy $E = 2\times10^{51}$\,erg, and the preSN radius of 550\rsun.
In this model the core of the CS shell is of 0.14\msun\ and the halo density 
 parameter is $A = 3.2\times10^{16}$\gcm.
In the second model (Figure 3) the CS core is absent and the density 
  through the CS shell is 
 $\rho = Ar^{-2}$ with $A = 3.8\times10^{16}$\gcm. 
Both models provide a comparable fit to the light curve and the expansion velocity.
The existence of the CS core and, therefore, of the heavy mass loss 
  related to the powerful preSN luminosity do not affect the light curve significantly.
Note that for the power law of the density distribution in the outer ejecta  $\rho \propto 1/v^n$ 
 the effect of the CS interaction turns out unchanged, if $M$ and $E$ obey the relation 
 $E \propto M^{(n-5)/(n-3)}$.

The luminosity maximum in the range of $0 \leq t < 27$ days is related to the CS interaction.
At $t \geq 27$ days the forward shock becomes adiabatic and its contribution to the bolometric 
  luminosity decreases significantly.
Note that the full width of the initial luminosity maximum essentially depends on the 
  radius of the CS shell halo, which permits us to find the value of $r_2$.


\begin{figure}
	\centering
	\includegraphics[trim = 50 120 0 0, width=0.97\textwidth]{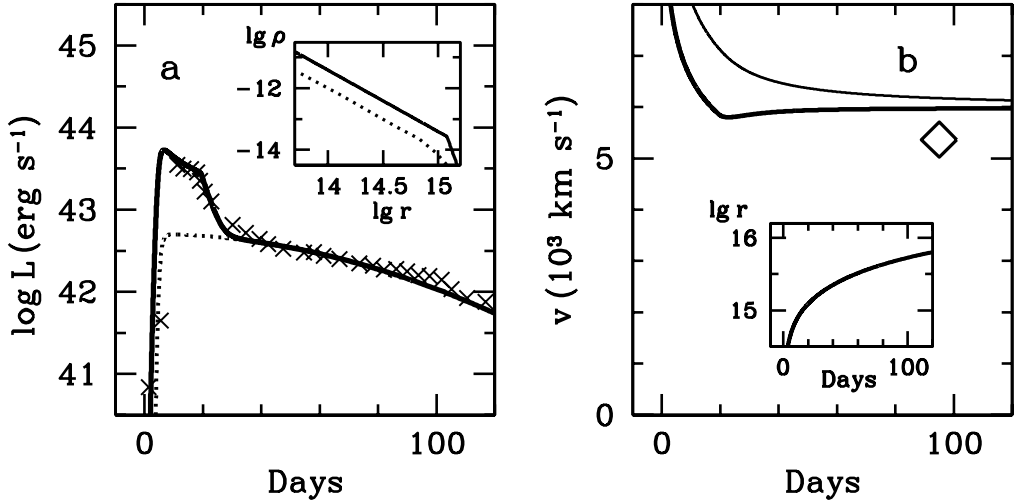}
	\caption{\rm
		The same as Figure 2, but for the CS density distribution $\rho \propto r^{-2}$.
		This model is also consistent with the observational constraints.
	}
	\label{fig:blc1}
\end{figure}


\subsection{Presupernova flash}

Modeling the hydrodynamic effect of the energy deposition in the central zone of the preSN 
  for a broad range of the mass coordinate $m_{dep}$ and the flash energy $E_{dep}$ 
  shows that the ratio $q = E_{dep}/E_b$  should be in the range of 
 $q \sim 0.2... 0.3$ in order to account for the preSN luminosity at the level of 
  $10^{40}$\ergs\ during several hundreds days. 
This conclusion is illustrated by models m2a and m1.4 c $m_{dep} = 2$\msun\ and 
  $m_{dep} = 1.4$\msun, respectively (Table 3). 
Both models have the similar light curves and the comparable $q$ (0.23 and 0.255) and yet 
 the values of $E_{dep}$ differ by a factor of 10.
  
Model m2a, adopted as a standard one, describes the preSN emission (Figure 4), 
 if the energy release occurred around 390 days before the supernova explosion. 
If the collapse had not happened, the preSN would have shone at the 
 level of $\sim 10^{40}$\ergs\ for about 1000 days (Figure 4, inset).
The internal energy deposited in this case corresponds to the burning of 0.003\msun\ of 
  C/O mixture.
The ejected mass is $0.13$\msun\ with the r.m.s. velocity $v_{rms} = 45$\kms\ and 
  the maximal velocity $v_{max} = 70$\kms.
In 390 days the ejecta boundary with the preSN radius taken into account expands up to 
    $2.7\times10^{14}$\,cm.
This determines the radius of the core of the SN shell, which is taken into account in 
  the CS interaction model.  
  
Figure 4 also shows the light curve of the model m2b with the deposited energy twice as lower. 
In this case there is no mass ejection and the shock energy brings about only pulsations with 
 the quasi-period of hundreds days and the average luminosity comparable to the equilibrium 
 luminosity of RSG.
Yet one cannot rule out that the increase of the internal energy of the RSG envelope might 
 bring about the enhanced preSN mass loss rate. 
However, to confirm this conjecture, one need to use a more refined model of the external layers 
where acoustic waves transform into weak shocks with the cooling and the dust formation.
That kind of modeling is beyond goals of the present paper.

 \begin{table}[t]
 	\vspace{6mm}
 	\centering
 	{{\bf Table 3} Model parameters of preSN flash. The mass in solar 
 		 units, and the energy in ergs; in parentheses is the power of ten.
  }
 	
 	\vspace{5mm}\begin{tabular}{l|c|c|c|c|c|c} 
 		\hline
 	Model  &  $m_{dep}$  &  $E_b$  & $E_{dep}$ & 
 	$M_{ej}$   & $E_{kin}$ & $q$ \\ 
 		\hline
 		&            &                   &        &   &   &   \\
 	
 m2a    & 2.0  & 2.4(49)   & 5.5(48)  & 0.124  & 2.5(45)  & 0.23 \\
 m2b   & 2.0  & 2.4(49)  & 2.75(48)   & 0      & 0         & 0.115 \\
 m1.4  & 1.4  & 2.1(50)  & 5.3(49)   & 0.236   & 5.5(49)   & 0.255 \\
 		\hline
 	\end{tabular}
 \end{table}

\begin{figure}
	\centering
	\includegraphics[trim = 50 120 0 30, width=0.97\textwidth]{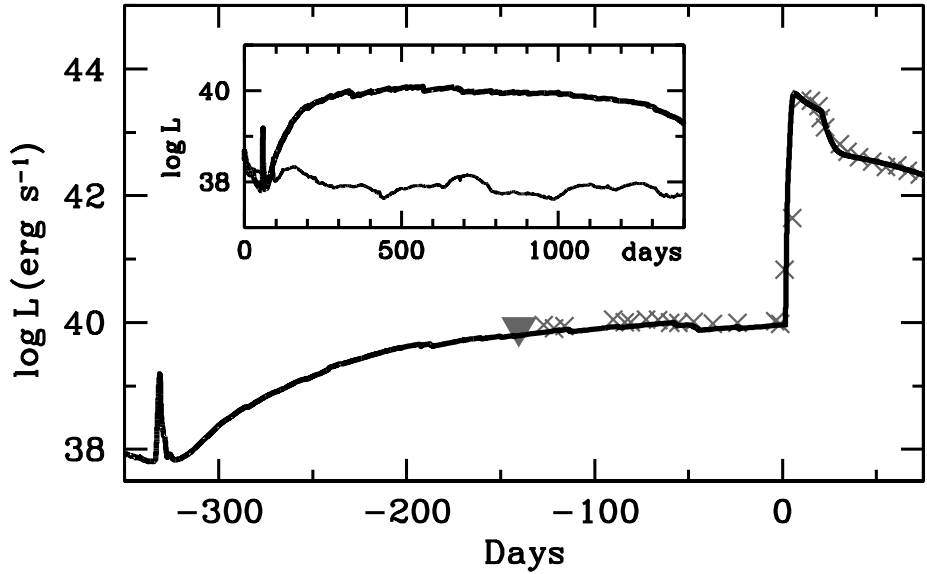}
	\caption{\rm
		Bolometric light curve of the preSN for the model m2a 
		(Table 3) combined with the model supernova light curve is compared to observational 
		data ({\em triangular symbol\/} is the 
		upper limit). The moment of the energy deposition 
		is 390 days before the collapse. 
		{\em Insert} shows the light curves of the preSN for the model 
		m2a ({\em thick line})	and for the model m2b ({\em thin line}) (Table 3), 
		which demonstrate strong dependence of 
		the observational effect on the deposited energy.
	}
	\label{fig:blc2}
\end{figure}

\section{Discussion}

Analyzing and modeling the photometric and spectral data on SN~2020tlf lead us to the picture 
  of the confined dense CS shell (Table 2) that had expanded before the explosion at 50\kms\ and 
  then has been accelerated by the supernova radiation up to 320\kms\ before the shock wave 
  by day 10. 
This shell had the outer radius of $\sim$10$^{15}$\,cm in line with the previous estimate 
 of Jacobson-Gal\'{a}n et al. (2022) and formed during $\sim$6 years before the explosion.
The spectral and photometric data do not show apparent signs of the material lost during 
 the stage of the powerful preSN luminosity.
However, hydrodynamic modeling of the high preSN luminosity as an outcome of the nuclear burning flash 
  $\sim$400\,days before the collapse predicts the ejection of $\sim$0.1\msun\ with the velocity 
  of $\approx$50\kms.
This material was present before the explosion within the radius of  $3\times10^{14}$\,cm and has 
been swept up by the supernova in 2 days after the explosion. 

A possibility of the nuclear burning flash follows from the vigorous convection in the O and Ne burning zone 
 at the final evolution of a massive preSN (Arnett \& Meakin 2011, Moc\'{a}k et al. 2018).
Current three-dimensional simulations of convective nuclear burning employ  a numerical domain
  bounded by rather narrow spherical angle, which is not able to treat a large-scale convection that 
  unavoidably should emerge (Chandrasekhar 1961). 
The large-scale convection results in the expansion of the burning zone and admixing fresh C/O material from 
  upper layers (Arnett \& Meakin 2011).
This can produce the powerful nuclear burning flashes accompanied by the generation of acoustic waves 
  that transform into shock waves.  

Our hydrodynamic modeling of the flash effects revealed two important facts: (i) the preSN luminosity at the level 
 of $10^{40}$\ergs\ occurs for the ratio  $q = E_{dep}/E_b \sim 0.2-0.3$, i.e., for the deposited energy substantially 
  lower than the binding energy; (ii) the preSN luminosity steeply drops for lower $q$.
The latter can explain why a phenomenon of the powerful preSN luminosity is so rare among SNe~IIP.  
As noted above (Section 3.2), the significant luminosity drop for  $q < 0.2$ nevertheless could be accompanied 
  by a strong mass loss.
This regime of the mass loss presumably formed the halo of the CS shell in SN~2020tlf.
Less massive, but dense confined CS shells observed in other supernovae, viz., SN~2013fs (Yaron et al. 2017) and 
 SN~1998S (Chugai 2001), are probably of the same origin as the halo of the SN~2020tlf CS shell.

The enhanced preSN mass loss via pumping the acoustic waves into the RSG envelope (Quataert \& Shiode 2012) includes the 
  excitation of gravitational hydrodynamic waves in the convective nuclear burning zone. 
Note in this regard that the turbulent convection in the burning zone is able to directly generate acoustic waves 
  via Lighthill (1952) mechanism.
The modification of the Lighthill mechanism for the stratified atmosphere (Stein 1967) provides the acoustic 
luminosity  
\begin{equation} 
L_w \approx \epsilon l^3  = C\rho c_s^3l^2\mathcal{M}^8 \sim
 2.3\times10^{43}\mathcal{M}^8_{0.1} \quad \mbox{erg}\,\mbox{s}^{-1},
\end{equation}
where the following values are used for the estimate: the density $\rho \sim 10^6$\gcmq, the depth of 
convective burning zone 
 $l\sim r \sim 5\times10^8$\,cm, the sound speed $c_s \sim 3\times10^8$\cms, Mach number for convective motion 
  $\mathcal{M} = v_{conv}/c_s = 0.1$. 
The factor $C$ is in the range of $180-450$ for Mach number in the range of $0.01-0.1$ (Stein 1967) and 
here is assumed to be 300.
At the Ne and O burning stage Mach number can reach $\sim$0.1 (Arnet \& Meakin 2011). 
With $\mathcal{M} = 0.07$ the acoustic luminosity is $\sim$10$^{42}$\ergs\ and during $\sim$10$^6$\,s the amount 
 of the acoustic energy  
  deposited in the RSG envelope can reach $\sim$10$^{48}$\,erg. 
This additional energy could significantly increase the mass-loss rate, but physics of this process requires  
 a clarification.
The strong dependence of the acoustic luminosity on $\mathcal{M}$ combined with the increase of the 
 burning rate with time can explain a significant growth of the mass loss rate at the final evolution stage, 
 year--decade prior to the core collapse.  

Table 4 demonstrates four CCSNe with the well-studied confined  circumstellar shell. 
The previous list of three supernovae (Chugai 2022) is enlarged with SN~2020tlf.
The table includes the supernova ID, type, the CS gas velocity, and the duration of the heavy mass loss. 
One can admit that the mass of the CS shell depends on the initial mass of the progenitor, which in turn 
 determines the burning rate and convection at the final year--decade. 
Noteworthy that SNe~IIP (SN~2013fs and SN~2020tlf) show significantly different mass of the CS shell 
 (0.003\msun\ vs. 0.37\msun), which presumably suggests that initial mass of SN~2013fs is lower 
 compared to SN~2020tlf.

The dependence of the phenomenon of a confined CS shell on the progenitor mass is emphasized 
  by the fact that two nearby CCSNe --- SN~1993J (type IIb) и SN~1987A (IIP) --- do not show 
  signatures of the confined dense CS shell in the early spectra.
For SN~1993J helium core mass inferred from hydrodynamic modeling of the light curve and expansion velocity 
 is $\approx$3\msun\ (Utrobin 1994), which suggests the progenitor mass of 12\msun\ (Woosley et al. 2002).
We conclude therefore that for the progenitor mass of $\sim$12\msun\ the conditions for the enhanced mass loss 
 year--decade before the core collapse are absent.
For SN~1987A the currently most confident estimate of the initial mass inferred in the framework of the model 
 that suggests merging of binary with initial stellar mass 15\msun\ + 7\msun\ at the stage of carbon core ignition 
  (Utrobin et al. 2021).
The merging affects neither carbon core formation, nor the subsequent core evolution.
The absence of the confined dense CS shell of SN~1987A therefore implies that in stars with the 
initial mass around 15\msun\ the mechanism of the enhanced mass loss year--decade before the 
 core collapse also does not operate.   
The unavoidable conclusion is that the confined dense shell is a feature of the preSN with the initial mass 
 $>$15\msun\ and, therefore, the initial masses of the progenitors of SN~2013fs and SN~2020tlf are larger than 15\msun.
 
The conjecture that the enhanced mass-loss rate year--decade before the explosion is a feature of massive stars is 
supported by  SN~2010al  (Table 4).
The preSN was a WR star of WN class (Chugai 2022) and therefore originated from a star with the initial mass of $\gtrsim$25\msun\ 
 (Woosley et al. 2002).
The large preSN mass of SN~2010al is indirectly supported by the low amount of $^{56}$Ni (if any). 
This suggests that almost all $^{56}$Ni experienced fallback, which is predicted for massive stars 
 $\gtrsim$25\msun\ (Woosley et al. 2002). 

If the dependence of the enhanced mass-loss rate for the preSN on the initial progenitor mass is the case,
 then the worrisome question arises, what is the actual progenitor mass for SN~2020tlf. 
This question remains open and then the modeling of the preSN flash assuming the initial mass of 12\msun\ 
 should be considered as an illustration of the underlying physics of the phenomenon that 
 is valid in the case of more massive preSN as well.
Note that the modeling of SN~2020tlf based on the radiation hydrodynamics is hampered by the absence 
  of spectra in the range of 10 -- 95 days, and, consequently, the expansion velocities required to constrain, 
  in combination with the light curve, the model parameters.

\begin{table}[t]
	\vspace{6mm}
	\centering
	{{\bf Table 4} Confined CS shell of CCSNe}
	
	\vspace{5mm}\begin{tabular}{l|c|c|c|c|c} 
		\hline
		SN name &	 Type       &  $M_{cs}$~($M_{\odot}$) & $u_{cs}$ (\kms) &  $^{56}$Ni ($M_{\odot}$) & $t_{cs}$ (yr)\\ 
		\hline
		&            &                   &        &   &    \\
		
		2013fs    & 		IIP    &    0.003     & 50    &   0.05 & 10 \\
		
		1998S     &		IIL   &      0.1     &   40      &    0.15 & 10\\
		
		2010al    &		Ibn         &     0.14    &  1000      & <0.015 & 0.4\\ 
		
		2020tlf    &		IIP      &     0.37$^{\bigstar}$   &  50    &   0.03 & 6 \\ 	
		\hline
	\end{tabular}
\parbox[]{11cm}{\small $^{\bigstar}$Combined mass of halo and core of the CS shell}
\end{table}

\section{Conclusion}

The light curve and spectra of SN~2020tlf indicate the presence of a confined dense shell with the radius 
 of $\sim$10$^{15}$\,cm that formed by the loss of $\approx$0.2\msun\  during $\sim$6 years before the explosion.
 
The powerful preSN luminosity is explained by the flash in the convective zone of nuclear burning 
 $\sim$400\,days before the collapse.
The shock wave initiated by the energy release results in the ejection of about 0.1\msun\ of the 
 preSN envelope with the average velocity of 50\kms. 
This material however did not manifest itself in observations after the supernova explosion, since it has been 
 swept up by the shock wave in two days after the explosion. 
 
\vspace{1cm} 

This study was supported by the Russian Scientific Foundation (RSF),
project number 19-12-00229, and the Russian Foundation for Basic Research
(RFBR) and the Deutsche Forschungsgemeinschaft (DFG), project number 21-52-12032.

\clearpage

\centerline{\bf References}
\bigskip

\begin{flushleft}
	
Arnett W. D., Astrophys. J. {\bf 237}, 541 (1980)\\
\medskip	
Arnett W. D., Meakin C., Astrophys. J. {\bf 733}, 78 (2011)\\	
\medskip
Chandrasekhar S. {\it Hydrodynamic and hydromagnetic stability}, Oxford: Clarendon, 1961\\
\medskip
Chevalier R. A., Astrophys. J. {\bf 259}, 302 (1982)\\
\medskip
Chugai N. N., eprint arXiv:2203.02717  (2022)\\
\medskip
Chugai N. N., Mon. Not. R. Astron. Soc. {\bf 326}, 1448 (2001)\\
\medskip
Dessart L., Livne E., Waldman R., Mon. Not. R. Astron. Soc. {\bf 405}, 2113 (2010)\\
\medskip
Fassia A., Meikle W. P. S., Vacca V. D. et al., Mon. Not. R. Astron. Soc. {\bf 318}, 1093 (2000) \\
\medskip
Giuliani J. L.,  Astrophys. J. {\bf 245}, 903 (1981) \\
\medskip
Hummer D. G., Mihalas D., Astrophys. J. {\bf 150}, L57 (1967)\\
\medskip 
Jacobson-Gal\'{a}n W. V., Dessart L., Jones D. O. et al., Astrophys. J. {\bf 924}, 15 (2022)\\	
\medskip 
Janka H.-T. {\it Handbook of Supernovae}, Springer (International Publishing AG, p. 1575, 2017) \\
\medskip	
Leung Shing-Chi, Wu S., Fuller J.,  Astrophys. J. {\bf 923}, 41L (2021)\\
\medskip
Lighthill M. J., Proc. Royal Soc. Series A  {\bf 211}, 564 (1952)\\
\medskip
Moc\'{a}k M., Meakin C., Campbell S. W.  et al., Mon. Not. R. Astron. Soc. {\bf 481}, 2918 (2018)\\
\medskip
Pastorello A., Benetti S., Brown P. J. et al., Mon. Not. R. Astron. Soc. {\bf 449}, 1921 (2015) \\
\medskip
Quataert E., Shiode J.,  Mon. Not. R. Astron. Soc. {\bf 423}, L92 (2012)\\
\medskip
Stein R. F., Solar Phys. {\bf 2}, 385 (1967)\\
\medskip 
Sunyaev R. A., Titarchuk L. G., Astron. Astrophys. {\bf 86}, 121 (1980)\\
\medskip
Sutherland R. S., Dopita M. A.,  Astrophys. J. Suppl. {\bf 88}, 253 (1993)\\
\medskip
Utrobin V. P., Wongwathanarat A., Janka H.-Th. et al., Astrophys. J. {\bf 914}, 4 (2021)\\
\medskip 
Utrobin V. P., Astron. Astrophys. {\bf 461}, 233 (2007)\\
\medskip 
Utrobin V. P., Astron. Astrophys. {\bf 281}, L89 (1994)\\
\medskip	
Woosley S. E., Heger A., Weaver T. A., Rev. Mod. Phys. {\bf 74}, 1015 (2002) \\
\medskip
Yaron O., Perley D. A., Gal-Yam A. et al., Nature Physics {\bf 13}, 510 (2017) \\

\end{flushleft}

\end{document}